\documentclass{ws-ijmpc}

\usepackage{ulem}
\usepackage{hyperref}
\usepackage{lipsum,verbatim}
\usepackage{graphicx}
\usepackage{multirow,rotate,subfigure}
\usepackage{amscd,amsmath,amssymb}
\usepackage{booktabs}

\usepackage{adjustbox}

\usepackage{mathrsfs}
\usepackage{mathtools}

\usepackage{color, colortbl}
\definecolor{Gray}{gray}{0.9} 
\definecolor{LightCyan}{rgb}{0.88,1,1}
\definecolor{ciano}{rgb}{0,1,1}
\definecolor{magenta}{rgb}{1,0,1}
\definecolor{amarelo}{rgb}{1,1,0}
\definecolor{iw}{rgb}{0.7,0.93,0.36}  
\newcommand{\iw}{\cellcolor{iw}}
\newcommand{\ia}{\cellcolor{amarelo}}
\newcommand{\im}{\cellcolor{ciano}}

\definecolor{cerise}{rgb}{0.96,0.0,0.63}

\usepackage{color}
\definecolor{red}{rgb}{1,0,0}
\definecolor{green}{rgb}{0,1,0}
\definecolor{blue}{rgb}{0,0,1}

\newcommand{\xxb}[1]{\textcolor{blue}{#1}}

\begin{document}

\markboth{Jason A.C.~Gallas}
{Field discriminants for cyclotomic period equations}

\catchline{}{}{}{}{}

\title{Field discriminants of cyclotomic period equations}

\vspace{-0.25truecm}

\author{Jason A.C.~Gallas} 

\address{
Instituto de Altos Estudos da Para\'\i ba,
  Rua Silvino Lopes 419-2502,\\   
  58039-190 Jo\~ao Pessoa, Brazil,\\
Complexity Sciences Center, 9225 Collins Ave.~1208, 
Surfside FL 33154, USA,\\
  Max-Planck-Institut f\"ur Physik komplexer Systeme,
  01187 Dresden, Germany\\
  jgallas@pks.mpg.de}

\maketitle

\begin{history}
\received{2 October 2019}
\accepted{4 November 2019}
\centerline{https://doi.org/10.1142/S0129183120500217}
\end{history}


\begin{abstract}
We show that several orbital equations and orbital clusters of the quadratic
(logistic) map coincide surprisingly with cyclotomic
{\it period equations}, polynomials whose roots are Gaussian periods.
An analytical expression for the field discriminant of period equations
is obtained and applied to discover and to fill gaps in number field databases
constructed by numerical search processes.
Such expression allows easy assess to inessential divisors of conventional
discriminants and sheds light into why numerical construction of databases
is a hard problem. It also provides significant information about
the organization of periodic orbits of the quadratic map.
\keywords{Quadratic dynamics; Symbolic computation;   Algebraic dynamics.}
\end{abstract}

\ccode{PACS Nos.:
      02.70.Wz, 
      02.10.De, 
      03.65.Fd} 

\section{Introduction}

It was recently discovered that, in the partition generating limit of the
celebrated quadratic (logistic) map $x_{t+1}=2-x_t^2$, the familiar pair of
period-3 equations of motion for trajectories in phase space can be extracted in
a surprisingly double manner from either one of the following 
{\it orbital carriers}\cite{g19}:
\begin{eqnarray*}
  \varphi_1(x) =\ \varphi_1(x;\sigma)
    &=& x^3 -\sigma\,x^2 -(\sigma^2-2\sigma+3)\,x 
   + \sigma^3-2\sigma^2+3\sigma-1,     \label{c1}\\
 \varphi_2(x) =\ \varphi_2(x;\sigma) 
 &=& x^3 -(1-\sigma)\,x^2 -(\sigma^2+2)\,x
  -\sigma^3+\sigma^2-2\sigma+1.        \label{c2}
\end{eqnarray*}
From  them, choosing either $\sigma=0$ or $\sigma=1$, one
obtains the period-$3$ orbits:
\begin{alignat}{5}
 \Phi(x) &=& \ \ \varphi_1(x;0) \  &=\  \varphi_2(x;1) 
  \  &=&\ \ x^3 -3x-1,  &\Delta_{\Phi(x)}  &=9^2,\label{fi}\\
                               \smallskip
\overline{\Phi}(x) &=& \ \ \varphi_1(x;1) \  &=\ \varphi_2(x;0) 
  \  &=&\ \ x^3 -x^2 -2x +1, \qquad  &\Delta_{\overline{\Phi}(x)} &=7^2.\label{ficonj}
\end{alignat}
Clearly, such observable {\it macroscopic}  orbits of the quadratic map are obtained as
independent ``projections'' from degenerate non-observable {\it microscopic} carriers,
either $\varphi_1(x)$ or $\varphi_2(x)$.
Thus, rather than independent orbits, $\Phi(x)$ and $\overline{\Phi}(x)$
are seen to emerge as a dual pair of conjugated orbits.
So,  in classical dynamics, multiple microscopic carriers define
a structural framework, or skeleton, for the orbits observed
macroscopically in phase space, a mind-boggling fact.
For details, see Ref.~1.

While working to determine orbital carriers for period-5 motions, we 
realized that the orbit $\overline{\Phi}(x)$ in Eq.~(\ref{ficonj})
is one of the  so-called {\it period equations} (defined in the next Section),
which are the key players introduced
in 1801 by Gauss to solve algebraically (i.e.~by finite chains of
radicals) the {\it cyclotomic equations}\cite{da,bach,reu,cox}.
The orbit $\overline{\Phi}(x)$ has the lowest possible discriminant
for cyclic cubics, namely $\Delta_{\overline{\Phi}(x)} =7^2$.
Its dual orbit, ${\Phi}(x)$, has the second smallest discriminant,
viz.~$\Delta_{\Phi(x)}  =9^2$.
These remarkable facts sparked our interest in period equations.
After computing many such equations, we found that they indeed
coincide with several orbital equations and clusters
of periodic trajectories of the quadratic map.
This unexpected coincidence makes period equations an object
of interest in the investigation of the dynamics of classical systems.

The purpose of this paper is to present a number of properties discovered
for families of period equations.
A key finding explored here in several applications
boils down to Eq.~(\ref{conjecture}) below,
an exact expression for the field discriminant of period equations.
Such expression reflects the fact, observed in our extensive
computations, that field discriminants of period equations were found to
be invariably given by powers of single prime numbers,
associated in a simple manner with the equations.

Our motivation is related to applications in physics and, accordingly,
we study equations with transitive cyclic groups, generically labeled
as $m$T1,  $m>1$.
Details concerning such applications may be found, e.g., in
Refs.~\cite{g19,res} and references therein.
An additional long-term and more ambitious goal is to search for 
a systematic way to interconnect arithmetically families of equations
of motion involved
in period-doubling cascades of the map, forming towers of
equations-within-equations of ever growing degrees.

In the remainder, we start by presenting briefly the context and
a few concepts needed to understand and to define Eq.~(\ref{conjecture}).
Then, we show
Eq.~(\ref{conjecture}) to be an expeditious tool for the computation of the so-called
{\sl inessential discriminant divisors}\cite{eg17,ha64}, factors 
of the conventional discriminant of equations of motion.
Finally, we present tables and several explicit expressions
for period equations and periodic orbits for relatively high periods 
of the quadratic map.
Our tables provide a good feeling for the distribution and growth of the
corresponding field discriminants.
To start, one first needs to come to grips with period equations.

\section{What are  period equations?}
\label{sec:what}

Period equations are auxiliary polynomials with integer coefficients
introduced by Gauss in his
book {\sl Disquisitiones Arithmetic\ae}\cite{da}, in the quest to solve
cyclotomic equations with ({\it finite\/} chains of) radicals, i.e.~to solve 
equations for the division of the perimeter of a circle into a
given number of equal parts\cite{bach,reu,cox}.
Since the methodology to obtain period equations is well described by Gauss,
it has been thought unnecessary to give the arithmetical details in full 
but just a brief summary. A detailed textbook exposition is given in a
classic book by Bachmann\cite{bach}.
An enthusiastic opinion and summary of Bachmann's  book is given by
Dedekind\cite{dede}.

Consider the set $\Omega$ of the $p-1$ complex roots of the equation $x^p-1=0$
where, here and throughout the paper, $p$ is a prime number.
To obtain Gauss'  period equations one starts by first distributing the
$p-1$ roots in $\Omega$, into certain sums called ``periods'', as demonstrated
in the {\sl Disquisitiones Arithmetic\ae}\cite{da}, in article 343:
{\it Omnes radices $\Omega$ in certas classes (periodos) distribuuntur}.
For two numbers $e$ and $f$ such that $p-1=ef$, Gauss partitions all roots in
$\Omega$ into $e$ disjoint classes, thereby forming $e$ periods $\eta_i$,
each one consisting of a sum of $f$ roots.

Let $r$ be any complex root in $\Omega$, for example $r=e^{2\pi i/p}$,
and $g$ a primitive root modulo $p$.
Then, the ``periods'' (not the period equations) are given by the sums:
\begin{equation}
  \eta_i = \sum_{k=0}^{f-1} r^{g^{ke+i}}, \qquad i=0,1,\cdots,e-1,
\end{equation}
or, more explicitly, $e$ sums of $f$ distinct complex roots suitably
selected from $\Omega$
\begin{alignat*}{8}
  \eta_0 &=& r &+&r^{g^e} &+&r^{g^{2e}} &+&\ \cdots\  &+&r^{g^{(f-1)e}},\\
  \eta_1 &=& r^{g} &+&r^{g^{e+1}} &+& r^{g^{2e+1}} &+&\ \cdots\ &+&\ r^{g^{(f-1)e+1}},\\
  &\; \; \vdots&\\
  \eta_{e-1} &=&\ r^{g^{e-1}} &+&\ r^{g^{2me1}} &+&\ r^{g^{3e-1}} &+&\ \cdots\
                  &+&\ r^{g^{fe-1}}.
\end{alignat*}    
The {\sl period equation} $\psi_e(x)$ of degree $e$ associated with $p=ef+1$
is defined as\cite{da,bach}
\begin{equation}
  \psi_e(x) = \prod_{k=0}^{e-1} (x-\eta_{k})
            = x^e+x^{e-1}+\alpha_2x^{e-2}+\cdots+ \alpha_e,    \label{theta}
\end{equation}
where all coefficients $\alpha_\ell$ turn out to be integers (from $\mathbb Z$).
Clearly, by construction the roots of  period equations are simply
the ``periods'' $\eta_i$ of Gauss.

To get a  feeling for the characteristic distribution of $p-1$ into products $ef$,
Table \ref{tab:tab01} illustrates the abundance and fast growth of the number
of roots in $\eta_i$.
For instance, the periods corresponding to the 50th prime of the form $p-1=2f$
contain $(233-1)/2=116$ roots in the summation,
the 500th prime contains $(3581-1)/2=1790$ roots,
and the 10000th prime contains $(104743-1)/2=52371$ roots.
In contrast,
for $e=23$ the corresponding number of roots are, respectively, $380, 5096$
and $132020$, an increase by factors of the order of  $2.5$.

\begin{table}[!tbh]
  \tbl{Primes $p=ef+1$ as a function of $e$,
    defining the degree of the period equations.
  Primes are far more distant from each other for prime $e$ 
  because  they have just a single signature (see text). 
}
{\begin{tabular}{@{}|r||r|r|r|r|r|r|r|r|@{}} \toprule        
    $e$ & $50$th & $200$th &$500$th & $1000$th &$2000$th & $3000$th & $5000$th & $10000$th\\
\midrule
 $ 2$& $ 233$& $ 1229$& $ 3581$& $ 7927$& $ 17393$& $ 48619$& $ 48619$& $ 104743$\\
 $ 3$& $ 577$& $ 2803$& $ 8089$& $ 17539$& $ 38011$& $ 104953$& $ 104953$& $ 225217$\\
 $ 4$& $ 577$& $ 2797$& $ 8009$& $ 17657$& $ 38153$& $ 105269$& $ 105269$& $ 225217$\\
 $ 5$& $ 1231$& $ 6151$& $ 17551$& $ 38231$& $ 82421$& $ 225781$& $ 225781$& $ 479821$\\
 $ 6$& $ 577$& $ 2803$& $ 8089$& $ 17539$& $ 38011$& $ 104953$& $ 104953$& $ 225217$\\
 $ 7$& $ 2143$& $ 9829$& $ 28057$& $ 59627$& $ 127709$& $ 351121$& $ 351121$& $ 748987$\\
 $ 8$& $ 1321$& $ 6521$& $ 17761$& $ 38609$& $ 82793$& $ 226241$& $ 226241$& $ 482441$\\
 $ 9$& $ 2089$& $ 9829$& $ 27361$& $ 59833$& $ 128467$& $ 351361$& $ 351361$& $ 747991$\\
 $ 10$& $ 1231$& $ 6151$& $ 17551$& $ 38231$& $ 82421$& $ 225781$& $ 225781$& $ 479821$\\
 $ 11$& $ 3433$& $ 17491$& $ 49369$& $ 105733$& $ 225611$& $ 610721$& $ 610721$& $ 1301851$\\
 $ 12$& $ 1321$& $ 6529$& $ 17989$& $ 38569$& $ 83101$& $ 226453$& $ 226453$& $ 481909$\\
 $ 13$& $ 4759$& $ 21191$& $ 59359$& $ 128519$& $ 275549$& $ 746227$& $ 746227$& $ 1592683$\\
 $ 14$& $ 2143$& $ 9829$& $ 28057$& $ 59627$& $ 127709$& $ 351121$& $ 351121$& $ 748987$\\
 $ 15$& $ 2971$& $ 14221$& $ 38671$& $ 83311$& $ 177601$& $ 481651$& $ 481651$& $ 1020961$\\
 $ 16$& $ 3041$& $ 14081$& $ 38321$& $ 82913$& $ 177409$& $ 481697$& $ 481697$& $ 1024337$\\
 $ 17$& $ 6257$& $ 28867$& $ 80173$& $ 177379$& $ 376687$& $ 1024523$& $ 1024523$& $ 2157709$\\
 $ 18$& $ 2089$& $ 9829$& $ 27361$& $ 59833$& $ 128467$& $ 351361$& $ 351361$& $ 747991$\\
 $ 19$& $ 7867$& $ 34961$& $ 95153$& $ 203339$& $ 434303$& $ 1164511$& $ 1164511$& $ 2460881$\\
 $ 20$& $ 3041$& $ 14081$& $ 39041$& $ 83621$& $ 176461$& $ 481181$& $ 481181$& $ 1025161$\\
 $ 21$& $ 4789$& $ 22051$& $ 60271$& $ 128941$& $ 277747$& $ 751549$& $ 751549$& $ 1588819$\\
 $ 22$& $ 3433$& $ 17491$& $ 49369$& $ 105733$& $ 225611$& $ 610721$& $ 610721$& $ 1301851$\\
 $ 23$& $ 8741$& $ 45403$& $ 117209$& $ 249229$& $ 533831$& $ 1443389$& $ 1443389$& $ 3036461$\\
\midrule
\end{tabular}}\label{tab:tab01}
\end{table}

\section{The dynamics behind period equations}

With hindsight,
it is not difficult to recognize now that wide classes of polynomial
equations of motion generated by a discrete-time physical model coincide with
equations considered by Gauss and by Abel in the early nineteenth century
  
Gauss introduced the systematic procedure reproduced in Section ({\ref{sec:what})
and used it to investigate the subgroups of the group, subsequently named
{\sl Galois group}, of the cyclotomic equations.
He found an explicit algorithm to solve a family of polynomials.
However, although interesting for many reasons, cyclotomic equations form
a relatively restricted class of polynomials.

Abel discovered that cyclotomic equations are nothing else than just a particular
case of a much wider class of equations\cite{abe}.
If the roots of an equation of arbitrary degree are connected in such a way that
{\it all roots} may be rationally expressed as a function of one of them,
say $x$ and, designating by $\theta(x)$ and $\theta_1(x)$ any two other roots,
where $\theta(x)$ and $\theta_1(x)$ are suitable rational functions,
one finds that they obey (the commutative composition law)
\begin{equation}
  \theta(\theta_1(x)) = \theta_1(\theta(x)),  \label{comut}
\end{equation}
then the equation in question will be always algebraically solvable by
finite chains of radicals.
Such general equations, called {\sl Abelian equations} after Kronecker,
are equations with $n$ roots $x_1, x_2, \cdots, x_n$ which satisfy the relations
\begin{equation}
  x_2=\theta(x_1),    \quad  x_3=\theta(x_2), \quad \cdots,\quad
  x_n=\theta(x_{n-1}), \quad  x_1=\theta(x_n), \label{abel}
\end{equation}  
where $\theta(x)$ is a rational function of $x$.
As observed by  Kronecker\cite{k53}, these general Abelian equations
are essentially cyclotomic equations, closing the cycle.

Anyone familiar with equations of motions of discrete-time dynamical systems will
immediately recognize that Eq.~(\ref{abel})  corresponds to the discrete-time
system
\begin{equation}
    x_{t+1} = \theta(x_t).  \label{dynsys}
\end{equation}
For algebraic functions $\theta(x)$, iterating this equation generates
sequences of polynomials whose roots are orbital points of the physical system
described by $\theta(x)$.
In general, orbital points are obtained numerically, i.e.~only in an
approximate way.
However, by studying equations of motion exactly the emphasis is shifted from
approximate {\sl orbital points} in phase space, to the study of exact analytical
properties and interrelations between equations of motion\cite{g19}.
Clearly, chaotic dynamics cannot be described by Abelian equations.
In the simple example chosen here, the partition generating limit,
all orbits are unstable.
However, carriers are valid generically, for arbitrary values
  of $a$. See Eq.~(11) in Ref.~1.

For applications in physics,
it is of interest to mention that in algebraic number theory it can be shown that
every cyclotomic field is an Abelian extension of the rational numbers $\mathbb Q$.
In this context, an important discovery is the so-called Kronecker-Weber theorem,
stating that every finite Abelian extension of $\mathbb Q$ can be generated
by roots
of unity, i.e.~Abelian extensions are contained within some cyclotomic field.
Equivalently, every algebraic integer whose Galois group is Abelian
can be expressed as a sum of roots of unity with rational coefficients.
For details see, e.g., Edwards\cite{edw}.
The key difficulty for application of the theorem above is buried in the
word ``some'': to find explicit algorithms providing effective  bridges
to implement the postulated interconnections between Abelian extensions
and cyclotomic fields.
After proving that wealth exists, it seems important to find ways to get to it.

\section{Results}

\subsection{Expression for the field discriminant of period equations}

Two known  invariants of any polynomial are its conventional discriminant $D$ and
the discriminant $\Delta$ of the number field $K$ underlying the
polynomial\cite{eg17,ha64,hc78}.
More technically,
let $p(x)$ be a monic  irreducible polynomial in $\mathbb Z(x)$
(i.e.~an irreducible polynomial over the integers with nonzero coefficient
of highest degree equal to 1), and $r$ a root of $p(x)\in \mathbb C$.
In addition, let $K$ be the number field $\mathbb Q(r)$ and $\mathcal O$
the ring of (algebraic) integers in $K$.
Then, the invariants $D$ and $\Delta$ are interconnected
by the simple-looking relation\cite{eg17,ha64,hc78}
\begin{equation}
  D=k^2 \Delta   \label{discri}
\end{equation}  
for some $k\in\mathbb Z$, where $D$ is the discriminant of $r$ and $\Delta$ is the
discriminant of  $\mathcal O$.
Again, the trouble lies in the word ``some''.

As pointed out by Vaughan\cite{v85},
``{\sl while $D$ can be found by straightforward (if tedious) computation,
the value of $k$ is quite another story.
According to Cohn\cite{hc78}, page 77, for
example, to determine $k$, one would have to test a finite number (which
may be very large) of elements of $K$ to see if they are integral.}''

Surprisingly, period equations form a wide class of equations for which
the computation of $\Delta$ and $k$ presents no difficulties and can be
done using the following wide-ranging result.

For any prime $p=ef+1$, the field discriminant $\Delta_e$ of the period equation
$\psi_e(x)$  in Eq.~(\ref{theta}) is given by
\begin{equation}\label{conjecture}
    \Delta_e =
    \begin{cases*}
  -p^{e-1}, & {\rm if} $(e-1) \hbox{ \rm mod } 4 =1 {\ \ \ \rm and\ \ \ }
                       f {\ \rm mod\ } 2 = 1 $, \\
  \phantom{-}p^{e-1}, & {\rm if otherwise}.
    \end{cases*}
\end{equation}
Equivalently, Eq.~(\ref{conjecture}) may be also written as
\begin{equation}\label{conj2}
    \Delta_e =
    \begin{cases*}
      (-1)^{n_P} p^{e-1}, & {\rm if} $(e-1) \hbox{ \rm mod } 4 =1$, \\
      p^{e-1},           & {\rm if otherwise},
    \end{cases*}
\end{equation}
where $n_P$ is the number of {\it pairs} of complex roots of $\psi_e(x)$.

The {\sl signature}\cite{ha64} of a polynomial is $(n_R, n_P)$,
sometimes written more economically as $n_R$,
where $n_R$ is the number of real roots of $\psi_e(x)$.
In the literature,  number field tables are normally ordered using
the magnitude $\vert\Delta_e\vert$ instead of $\Delta_e$.
Surprisingly, in Eq.~(\ref{conj2}) the sign of $\Delta_e$ is found 
to depend explicitly on the the nature, odd or even, of the total
number of pairs of complex roots of $\psi_e(x)$.
Therefore, we expect the determination of this sign to be a
non-trivial theoretical problem.

Equations (\ref{conjecture}) and (\ref{conj2}) are empirical expressions
distilled by consolidating numerical evidence gathered by tabulating thousands
of field discriminants for period equations for primes $p= ef+1$,
for $f$ varying up to a few thousands when $e\leq10$, and
for $f$ varying up to a few  hundreds when $11\leq e\leq60$.
Beyond $e=60$ computations become too sluggish and were not further
pursued. Despite the vast literature on cyclotomic equations,
we have not been able to locate Eqs.~(\ref{conjecture}) and (\ref{conj2}).
They correctly reproduce all numerically computed discriminants for the
total mass of data investigated.

For every prime $p=6f+1$, D.H.~Lehmer and E.~Lehmer \cite{ll84} reported
coefficients for $\psi_e(x)$ in terms of $L$ and $M$ in the quadratic partition
$4p=L^2+27M^2$. They also reported an explicit formula for the conventional
discriminant $D$ of $\psi_e(x)$.
Four explicit examples of $\psi_e(x)$ and discriminants were given.
However, while their $\psi_e(x)$ and the magnitudes of the conventional discriminants
are correct, we find the sign of all their discriminants to be incorrect.
In any case, nowadays it seems considerably safer and much easier to generate
$\psi_e(x)$ numerically than to use the quite long and intricate expressions
provided for the  coefficients of $\psi_e(x)$. 

For primes $p=8f+1$, E.~Lehmer\cite{l55} investigated the use of difference
sets and a class
of octic residues of $p$ to obtain conditions for octic period equations which,
according to her, ``{\sl are rather rare; there are only three such primes less
than ten thousand, namely $p=73, \ 6361$, and $9001$}''.
No explicit period equations were reported for these primes, only the
conventional discriminant $D$ for $p=73$, viz.~$D=2^{54}\cdot3^4\cdot73^7$.
For these rare octics, we find:
\begin{eqnarray*}
  \psi_8^{(73)}(x) &=& {x}^{8}+{x}^{7}+5\,{x}^{6}-17\,{x}^{5}-46\,{x}^{4}
            -136\,{x}^{3}+320\,{x}^{2}+512\,x+4096,\\
    &&\quad\qquad k^2=2^{54}\cdot 3^4, \qquad \Delta=73^7,\\
  \psi_8^{(6361)}(x) &=& {x}^{8}+{x}^{7}+398\,{x}^{6}+41446\,{x}^{5}-250747\,{x}^{4}
  +16689725\,{x}^{3}\cr
  &&\quad\ +486181868\,{x}^{2}-5601819268\,x+224934834784,\\
   &&\quad \qquad k^2=2^{90}\cdot 3^{12}\cdot5^{14}\cdot11^{12}, \qquad\Delta=6361^7,\\
  \psi_8^{(9001)}(x) &=& {x}^{8}+{x}^{7}+563\,{x}^{6}-42614\,{x}^{5}-556282\,{x}^{4},
  -28875030\,{x}^{3}\cr
   &&\quad\ +863797853\,{x}^{2}+13357557897\,x+926791611419,\\
   &&\quad \qquad k^2=2^{32}\cdot 3^{4}\cdot5^{14}\cdot11^{4}\cdot23^{12}
                   \cdot43^{12}, \qquad\Delta=9001^7.
\end{eqnarray*}  
They are the 3rd, 196th, and  271th octics for primes of the form $p=8f+1$,
respectively.
The discriminants for $p=73$ agree.
Although the reference table for totally complex octics lists polynomials containing
field discriminants up to 122 digits, $\psi_8^{(6361)}(x)$ and $\psi_8^{(9001)}(x)$,
with discriminants of 27 and 28 digits, respectively, are not listed.

Equation (\ref{conjecture}) gives a handy criterion to sort out equations
with either $k^2=1$ or $k^2\neq1$.
Thus, knowledge of field discriminants allows one to extract inessential
discriminant
divisors through a simple division of two (possibly very large) integers.

By avoiding the need for factorizing very large numbers,
Eq.~(\ref{conjecture}) allows a very significant reduction of the computations
required to assess the arithmetical scaffolding underlying period equations.

\subsection{A general expression for $\alpha_2$}
\label{sec:linear}

For  primes $p=3f+1$, a general period equation $\psi_3(x)$
solving the cyclotomic trisection problem was given in 1872
by Bachmann, on pages 210-213 and 224-230 of his classic book
{\it Die Lehre von der Kreisteilung}\cite{bach}, used properties of the
elementary symmetric functions $\eta_\ell$, to derive a one-parameter cubic
that we write as:
\begin{equation}
  \psi_3(x) = x^3 + x^2 -\tfrac{1}{3}(p-1)x + A,  \label{ep3}
\end{equation}  
An equivalent form having the same discriminants is $-\psi_3(-x)$.
Subsequently, in 1879, Cayley\cite{cay} reported 
\begin{equation}
  \psi_3(x) = x^3 + x^2 -\tfrac{1}{3}(p-1)x + fg-h^2,  \label{cay}
\end{equation}  
tabulating $f$, $g$ and $h$ for the 11 primes $p=3f+1$ below 100,
namely $7$, $13$, $19$, $31$, $37$, $43$, $61$, $67$, $73$, $79$ and $97$.
In 1901, Burnside showed\cite{bur} that Eq.~(\ref{cay})
``may be completely solved,
without the use of tables of any kind, by a number of trials which is small
in comparison with the prime considered.'' As an example, for $p=1213$,
with 5 trials he finds the correct solution $x^3+x^2 -404x+669=0.$

For $p=5f+1$, the period equation is a three-parameter quintic\cite{bu14}
\begin{equation}
  \psi_5(x) = x^5 + x^4 -\tfrac{2}{5}(p-1)x^3 + Cx^2 +Bx + A.  \label{ep5}
\end{equation}  

In general, for $e$ odd we find the coefficient $\alpha_2$ of the third
largest power of $x$ in $\psi_e(x)$, Eq.~(\ref{theta}), to be an integer
given by
\begin{equation}
 \alpha_2 = -\frac{e-1}{2e}(p-1) = -\tfrac{1}{2}(e-1)f.   \label{geral}
\end{equation}
We also find this same coefficient to be valid for $e$ even and
signature $(e,0)$. For $e=2$, the third largest power of $x$ in $\psi_e(x)$
is in fact the constant term of a quadratic.
We have not been able to locate this general coefficient in the literature.

It would be interesting to explore the possibility of, say, following Bachmann,
to use the elementary symmetric functions to obtain expressions for additional
coefficients.

\subsection{Tables of period equations}
\label{sec:tab}

All results reported here were obtained with a special purpose
MAPLE routine written to generate systematically large
sequences of period equations $\psi_e(x)$  for primes  $p=ef+1$, with
$e$  arbitrary but fixed. In this endeavor,
period equations tabulated in 1875 by Reuschle\cite{reu} were 
helpful to validate our routine.
Reuschle would be certainly amazed, perhaps shocked, to see that every
period equation recorded in his invaluable and influential work
of 13 years\cite{fn06}, could be now reproduced in fractions of a second
or just a few seconds.
For instance, the first 20 period equations for $p=20f+1$ were generated
in 3.7s, for $p=30f+1$ in 8.1s, for $p=40f+1$ in 17.2s,
for $p=50f+1$ in 29.3s, and for $p=60f+1$ in 83.9s, running
MAPLE 2014 on a modest and aging DELL XPS 13 Ubuntu notebook.
These simple tests generated much more period equations than reported
in Reuschle's book.

In the continuation, we compare our results with the ones in the detailed
number field database of  Kl\"uners and Malle\cite{klu,km01,km00},
taken to be our reference tables.
Malle\cite{mal} presents an impressive table listing the first 15 million cyclic
cubic fields, complete up to field discriminant $10^6$.
These works contain links to a number of additional papers and online tables.
For applications in physics, we mention the tables of totally real number fields
up to degree 10 computed and maintained by Voight\cite{voi}.
Most tables are concerned with number fields of relatively low-degree.
The database of Kl\"uner and Malle has minimal polynomials for fields up to
degree 19, degrees not available in tables known to us.
Of course, online number field tables are not at all concerned with
period equations and, accordingly, the majority of their polynomials
are not period equations. In fact, a byproduct of our work is precisely
to have identified an infinite family of polynomials responsible for producing
discriminants with the simplest possible structure, namely
powers of single prime numbers. 

For larger values of $e$, Table \ref{tab:tab05} below presents  information
that goes well beyond what is presently available for equations with
cyclic group.

\begin{table}[!tbh]
\tbl{Solution set for cubics with transitive group $3T1$, ordered by the value
of $p$, for the first 105 primes $p=3f+1$.
    Their field discriminant is  $\Delta=p^2$ and the
     minimal polynomial is $f(x)=x^3+x^2 -\tfrac{p-1}{3}x +A$.
  The {\sl inessential discriminant divisors} are defined by $k^2$.
  Solutions for highlighted primes are not in the reference tables.}
{\begin{tabular}{@{}|rrrr|rrrr|rrrr|@{}} \toprule
 \#&$p$ &$A$&$k$&   \#&$p$ &$A$&$k$&  \#&$p$ &$A$&$k$\\   
\midrule
 1& 7& $-1$& 1&            36& 379&$ 365$&5&           71& \iw853&\iw 1011&\iw$3^2$\\
 2& 13& $1$& 1&            37& 397&$-544$&$2^2$&       72& 859&$-509$&11\\
 3& 19& $-7$& 1&           38& 409&$-515$&5&           73& \iw877&\iw 1819&\iw 1\\
 4& 31&$-8$& 2&            39& 421&$-343$&7&           74& \iw883&\iw 1439&\iw 7\\
 5& 37&$11$& 1&            40& 433&$-16$&$2^3$&        75& 907&$-739$&11\\
 6& 43,&$8$&2&             41& 439&$-504$&$2\cdot3$&   76& \iw919&\iw$-1872$&\iw$2\cdot3$\\
 7& 61&$-9$& 3&            42& 457&$-220$&$2^3$&       77& \iw937&\iw$-2221$&\iw 1\\
 8& 67&$ 5$& 3&            43& 463&$ 343$&7&           78& \iw967&\iw 1361&\iw$3^2$\\
 9& 73& $-27$&3&           44& 487&$-505$&7&           79& \iw991&\iw$-2349$&\iw3\\
 10& 79&$ 41$& 1&          45& 499&$536$&$2\cdot3$&    80& 997&$-480$&$2^2\cdot3$\\
 11& 97&$ -79$& 1&         46& 523&$-891$&3&           81& \iw1009&\iw$-1719$&\iw$3^2$\\
 12& 103&$ -61$& 3&        47& 541&$ 521$&7&           82& 1021& 416&$2^2\cdot3$\\
 13& 109&$ -4$& $2^2$&     48& 547&$-81$&$3^2$&         83& \iw1033&\iw 1913&\iw7\\
 14& 127&$ 80$&2&          49& 571&$-719$&7&            84& \iw1039&\iw 2155&\iw5\\
 15& 139&$ 103$& 1&        50& 577&$ 171$&$3^2$&        85& \iw1051&\iw$-2608$&\iw2\\
 16& 151&$ -123$& 3&       51& 601&$512$&$2^3$&         86& \iw1063&\iw 2441&\iw 1\\
 17& 157&$ 64$& $2^2$&     52& \iw607&\iw$-1169$&\iw 1&       87& \iw1069&\iw 2336&\iw$2^2$\\
 18& 163&$ -169$& 1&       53& 613&$ 999$&3&            88& \iw1087&\iw$-2335$&\iw7\\
 19& 181&$-67$&5&          54& 619&$ 321$&$3^2$&   89& \iw1093&\iw$-1012$&\iw$2^2\cdot3$\\
 20& 193&$ 143$&3&         55& \iw631&\iw$-1075$&\iw5&         90& \iw1117&\iw 2565&\iw3\\
 21& 199&$ 59$&5&          56& \iw643&\iw$-1024$&\iw$2\cdot3$& 91& \iw1123&\iw 1331&\iw11\\
 22& 211&$ -125$&5&        57& \iw661&\iw$-1273$&\iw3&         92& \iw1129&\iw$-2927$&\iw 1\\
 23& 223&$ -256$&2&        58& 673&$-997$&7&           93& 1153&$-427$&13\\
 24& 229&$ -212$&$2^2$&    59& 691&$128$&$2\cdot5$&    94& \iw1171&\iw 347&\iw13\\
 25& 241&$ 125$&5&         60& \iw709&\iw$1313$&\iw 1&       95& \iw1201&\iw 2491&\iw7\\
 26& 271&$ 261$&3&         61& \iw727&\iw$1104$&\iw$2\cdot3$&   96& 1213& 629& 13\\
 27& 277&$ 236$&$2^2$&     62& \iw733&\iw$1276$&\iw$2^2$&        97& \iw1231&\iw$-1003$&\iw13\\
 28& 283&$ 304$&2&         63& 739&$-520$&$2\cdot5$&   98& \iw1237&\iw 1741&\iw11\\
 29& 307&$-216$&$2\cdot3$& 64& \iw751&\iw$1057$&\iw7&        99& \iw1249&\iw 2313&\iw$3^2$\\
 30& 313&$ 371$& 1&        65& 757& 729&$3^2$&         100& \iw1279&\iw$-2179$&\iw11\\
 31& 331&$-49$&7&          66& \iw769&\iw$-1481$&\iw5&       101& \iw1291&\iw$-3347$&\iw5\\
 32& 337&$ 25$&7&          67& \iw787&\iw$-991$&\iw$3^2$&    102& \iw1297&\iw$-1345$&\iw$13$\\
 33& 349&$ -517$& 1&       68& \iw811&\iw 1592&\iw2&         103& \iw1303&\iw$-2799$&\iw$3^2$\\
 34& 367&$ 435$&3&         69& 823& 61&11&            104& \iw1321&\iw 3327&\iw3\\
 35& 373&$ -221$&7&        70& 829&$-307$&11&         105& 1327& $-344$ & $2\cdot7$ \\
\midrule
\end{tabular}\label{tab:tab02}}
\end{table}

\begin{table}[!tbh]
\tbl{Solution set for biquadratics with field discriminant
$\Delta=p^3$ for primes $p=4f+1$,            
  and minimal polynomial $x^4+x^3+Cx^2 +Bx +A$, highlighted by
  signature.
The {\sl inessential discriminant divisors} are defined by $k^2$.
``Seq'' and ``Sig'' refer to the sequential enumeration of primes
and signature.
Among the first 80 primes there are 37 of signature 4 and 43 of signature 0.
Note that $C=-\tfrac{3}{8}(p-1)$ for totally real quartics.
}
{\begin{tabular}{@{}|c|c|r|c|r||c|c|r|c|r|@{}} \toprule
Seq & $p$ & C,B,A & Sig  & $k$ & Seq & $p$ & C,B,A & Sig  & $k$\\
\midrule
1,1& \ia5&\ia 1,1,1&\ia$0$ &$1$ & 
      41,18& 433& -162,839,-1003&$4$  &$2\cdot3^{3}$\\
 2,2& \ia13&\ia 2,-4,3&\ia$0$  &$3$ &
      42,19& 449& -168,-477,335&$4$  &$2\cdot5^{3}$\\ 
 3,1& 17& -6,-1,1&$4$  &$2$ & 
      43,20& 457& -171,1114,-2044&$4$  &$2$\\
 4,3& \ia29&\ia 4,20,23&\ia$0$  &$7$ &
      44,24& \ia461&\ia 58,-1066,4601&\ia$0$  &$5\cdot109$\\
 5,4& \ia37&\ia 5,7,49&\ia$0$  &$3\cdot7$ &
      45,25& \ia 509&\ia 64,350,8993&\ia$0$  &$11\cdot97$\\
 6,2& 41& -15,18,-4&$4$  &$2$ &  
      46,21& 521& -195,-814,-116&$4$  &$2\cdot5^{3}$\\
7,5& \ia53&\ia 7,-43,47&\ia$0$  &$13$ &
      47,26& \ia 541&\ia 68,1454,6921&\ia$0$  &$3\cdot5\cdot43$\\
8,6& \ia61&\ia 8,42,117&\ia$0$  &$3\cdot13$ &
      48,27& \ia 557&\ia 70,-1288,7439&\ia$0$  &$7\cdot127$\\
 9,3& 73& -27,-41,2&$4$  &$2^{4}$ &  
 49,22& 569& -213,818,-20&$4$  &$2\cdot5^{3}$\\
 10,4& 89& -33,39,8&$4$  &$2^{4}$ &
      50,23& 577& -216,-36,1296&$4$  &$2^{4}\cdot3^{3}$\\
11,5& 97& -36,91,-61&$4$  &$2$  &  
      51,24& 593& -222,-1816,-3968&$4$  &$2^{4}$\\
12,7& \ia101&\ia 13,19,361&\ia$0$  &$5\cdot19$ & 
      52,25& 601& -225,263,1256&$4$  &$2^{4}\cdot3^{3}$\\
13,8& \ia 109&\ia 14,-34,393&\ia$0$  &$3\cdot5\cdot7$ &
      53,28& \ia 613&\ia 77,1341,10773&\ia$0$  &$3^{2}\cdot7\cdot19$\\
14,6& 113& -42,-120,-64&$4$  &$2^{4}$ &
      54,26& 617& -231,-1581,-2374&$4$  &$2^{7}$\\
15,7& 137& -51,-214,-236&$4$  &$2$ & 
      55,27& 641& -240,1883,-4169&$4$  &$2$\\
16,9& \ia 149&\ia 19,-121,635&\ia$0$  &$5\cdot31$ &
      56,29& \ia 653&\ia 82,1102,13537&\ia$0$  &$7\cdot11\cdot19$\\
17,10& \ia 157&\ia 20,-206,517&\ia$0$  &$3\cdot37$ &
      57,30& \ia 661&\ia 83,2107,9427&\ia$0$  &$3\cdot163$\\
18,11& \ia 173&\ia 22,292,667&\ia$0$  &$43$ & 
       58,28& 673& -252,-2061,-4293&$4$  &$2\cdot3^{3}$\\
19,12& \ia 181&\ia 23,215,975&\ia$0$  &$3\cdot5\cdot13$ &
       59,31& \ia 677&\ia 85,127,16129&\ia$0$  &$13\cdot127$\\
20,8& 193& -72,-205,-49&$4$  &$2\cdot3^{3}$ &
       60,32& \ia 701&\ia 88,482,17117&\ia$0$  &$7\cdot13\cdot19$\\
21,13&\ia 197&\ia 25,37,1369&\ia$0$  &$7\cdot37$ &
        61,33& \ia 709&\ia 89,-1285,14853&\ia$0$  &$3\cdot7^{2}\cdot11$\\       
22,14&\ia  229&\ia 29,-415,933&\ia$0$  &$3\cdot19$ &
         62,34& \ia 733&\ia 92,-2428,9927&\ia$0$  &$3\cdot61$\\
23,9& 233& -87,335,-314&$4$  &$2^{4}$ &
         63,35& \ia 757&\ia 95,899,19407&\ia$0$  &$3\cdot7^{2}\cdot13$\\
24,10& 241& -90,-497,-739&$4$  &$2$ & 
        64,29& 761& -285,-1950,-2500&$4$  &$2\cdot5^{3}$\\
25,11& 257& -96,-16,256&$4$  &$2^{7}$ & 
    \im    65,30& 769& -288,2259,-4617&$4$  &$2\cdot3^{3}$\\
  \im   26,15& \ia269&\ia 34,454,1945&\ia$0$  &$5\cdot61$ &
        66,36&  \ia773&\ia 97,1691,17933&\ia$0$  &$11\cdot163$\\
27,16& \ia 277&\ia 35,329,2427&\ia$0$  &$3\cdot7\cdot19$ &
        67,37&  \ia797&\ia 100,-1046,20557&\ia$0$  &$13\cdot157$\\
28,12& 281& -105,123,236&$4$  &$2^{7}$ & 
        68,31&  809& -303,354,2348&$4$  &$2\cdot7^{3}$\\
29,17& \ia 293&\ia 37,641,1853&\ia$0$  &$73$ &
        69,38&\ia  821&\ia 103,2617,16327&\ia$0$  &$7\cdot193$\\
30,13& 313& -117,450,-324&$4$ &$2\cdot3^{3}$ &
        70,39& \ia 829&\ia 104,-2746,14025&\ia$0$  &$3\cdot5\cdot67$\\
31,18& \ia 317&\ia 40,-416,2827&\ia$0$  &$7\cdot67$ &
        71,40& \ia 853&\ia 107,-2399,17923&\ia$0$  &$3^{2}\cdot193$\\
 32,14& 337& -126,316,104&$4$  &$2^{7}$ &
        72,32&  857& -321,2946,-7636&$4$  &$2$\\
 33,19& \ia 349&\ia 44,240,4203&\ia$0$  &$3^{2}\cdot67$ &
        73,41& \ia 877&\ia 110,3234,16317&\ia$0$  &$3\cdot7\cdot31$\\
 34,15& 353& -132,684,-928&$4$  &$2^{4}$ & 
        74,33&  881& -330,2588,-4904&$4$  &$2^{7}$\\
 35,20&\ia  373&\ia 47,-303,4527&\ia$0$  &$3^{2}\cdot73$ & 
        75,34& 929& -348,-2845,-4997&$4$  &$2\cdot5^{3}$\\
 36,21&\ia  389&\ia 49,851,3773&\ia$0$  &$5\cdot7\cdot13$ & 
        76,35&  937& -351,-2401,-2434&$4$  &$2^{4}\cdot3^{3}$\\
 37,22&\ia  397&\ia 50,-918,3069&\ia$0$  &$3\cdot97$ &
         77,42& \ia 941&\ia 118,3470,19625&\ia$0$  &$5\cdot229$\\
 38,16& 401& -150,-25,625&$4$  &$2\cdot5^{3}$ & 
         78,36&  953& -357,1370,1396&$4$  &$2\cdot7^{3}$\\
 39,17& 409& -153,-230,548&$4$  &$2\cdot5^{3}$ & 
         79,37&  977& -366,-3969,-11911&$4$  &$2$\\
 40,23&\ia  421&\ia 53,-763,4557&\ia$0$  &$3\cdot7\cdot31$ &
 80,43&\ia  997&\ia 125,-3801,19017&\ia$0$  &$3\cdot13\cdot19$\\  
\midrule
\end{tabular}}\label{tab:tab03}
\end{table}

\subsubsection{Period equations for primes of the form $p=3f+1$}
\label{sec:ide}

The conventional polynomial discriminant of the cubic $\psi_3(x)$,
Eq.~(\ref{ep3}), is
\[ D_c = -27A^2-(6p-2)A +\tfrac{1}{27}(4p-1)(p-1)^2. \]
For $k=1$, Eq.~(\ref{conjecture}) implies $D_c=p^2$,
a quadratic in $A$ which has a rational and an integer value of $A$
as solutions. For $k\neq1$, both solutions are quadratic numbers.
Therefore, the constraint $D_c=p^2$ provides a simple an efficient
algorithm to sort out period equations with and without inessential
discriminant divisors.
For a fixed value of $e$, by increasing $f$ successively 
it is possible to extract in a systematic way, without omissions,
a list of all primes $p=ef+1$.
For example,
the 1187th cubic is ${x}^{3}+{x}^{2}-22569406\,x+41261890201$,
for $p=67708219$.
The 1631st cubic is ${x}^{3}+{x}^{2}-47112146\,x+124449351881$,
for $p=141336439$.
The 2405th cubic is ${x}^{3}+{x}^{2}-111367856\,x+452326735601$,
for $p=334103569$.
The (weak) growth of the number of cubics with $k=1$ obeys
a power law.

For cyclic fields, the reference table displays discriminants for
211 cubics,
the last one being $2989441=7^2\cdot13^2\cdot19^2$.
Among them, there are 66 discriminants of the form $p^2$, the
four ones larger than $1000$  being $1021, 1153, 1213$ and $1327$,
respectively the 82th, 93th, 96th, and 105th primes $p=3f+1$.
These 66 cases lead us to suspect that the corresponding
cubics could be  period equations, or isomorphic forms of
period equations. Indeed, they are.

Table \ref{tab:tab02} records $A$ and $k$ for the first 105 primes
$p=3f+1$,  characterized by field discriminants $p^2$.
The sign and magnitude of $A$ varies sensibly with $p$.
Highlighted primes are not listed in the reference table.
The magnitude of $A$ for all primes in the reference table is
smaller than $1000$.
With two exceptions, all missing  primes highlighted in Table \ref{tab:tab02}
have $\vert A\vert>1000$.
It is important to stress that the reference tables are not concerned with
period equations. They contain many discriminants which factor into
products of powers of several primes and have minimal polynomials
with coefficients that exceed $1000$ considerably.

\subsubsection{Period equations for primes  $p=4f+1$}

For cyclotomic primes $p=4f+1$ there are two classes of 4T1 cyclic polynomials,
characterized by signatures  $4$ or $0$.
Table \ref{tab:tab03} contains the first 80 period equations,
independently of signature.
Highlighting is used to discriminate signatures.

For $p=4f+1$, the reference tables\cite{klu} list 238 polynomials of
signature $4$ and 198 of signature $0$.
The largest cyclotomic prime listed is $p=769$ for signature $4$,
and $p=269$ for signature $0$.
Minimal polynomials for the totally real signature agree with ours,
modulo the trivial substitution  $x\mapsto -x$ or isomorphisms.
In contrast, for signature $(0,2)$, the reference table misses  primes
$109$, $149$, $157$, $173$, $181$,  $229$.

For $p=37$ and signature $0$, the minimal polynomial in the reference table is
\[ f(x) = x^4 + 2x^3 + 20x^2 + 19x + 7, \qquad k^2=3^4,\qquad \Delta_e=37^3, \]
while our Table \ref{tab:tab03} has
\[ g(x) = x^4 + x^3 +5x^2 +7x+49, \qquad k^2=3^2\cdot 7^2,\qquad \Delta_e=37^3. \]

\begin{table}[!tbh]
\tbl{Relative abundance of period equations as a function of the degree of their
    minimal polynomials and signature.
    $N$ refers to the number of fields listed in the reference tables.
    Highlighted blue boldface numbers are missing in the reference tables.
    See text for description of remaining data.
}
{\begin{tabular}{@{}|c|c|r|l|r|r|@{}}\toprule
Deg & Sig & $N$ & Discriminant bases & \#dig & Seq\\
\midrule
10 & 10 & 181 &$41, 61, 101, 181, \xxb{\bf 241,281,401,421,461,
      521,541,601} $  & 48 & $[2\cdot10^5]$\\ 
   &  0 &  79 &$-11, -31, -71, -131, -151, -191,
    \xxb{\bf -211,-251,-271}$
                    &22 &13\\
11 & 11 & 40& $23, 67, 89, \xxb{\bf 199,331,353,397,419,463,617,\cdots},
       920371$ &357 & $[10^{36}]$ \\
12 & 12 & 102& $73, \xxb{\bf 97,193, 241, 313,337,409,433,457,577,601,673}   $ &35 & 47\\
   &  0 & 91& $13, 37, 61, \xxb{\bf 109,157,181,229,277,349,373,397,421} $  &70 & $[2\cdot10^6]$\\
13 & 13 & 3 & $53,79,\xxb{\bf 131,157,313,443,521,547,599,677,859,911}$
      &27 & 4\\
14 & 14 & 75& $29, \xxb{\bf 113,197,281,337,421,449,617,673,701,757,953  }$ & 83& $[6\cdot10^6]$ \\
   &  0 & 56& $-43,\xxb{\bf -71,-127,-211,-239,-379,-463,-491,-547  }$ & 59& 517 \\
15 & 15 & 70& $31, 61,\xxb{\bf 151,181,211,241,271,331,421,541,571,601   }$ &83 & 7216\\
16 & 16 & 30& $\xxb{\bf97, 195, 257, 353,
      449, 577, 641, 673, 769,929,1153 }$&47 & 77  \\ 
   &  0 & 29& $17,\xxb{\bf 113,241,337,401,433,593,881,977,1009,1201} $  & 38& 7 \\
17 & 17 & 4& $103, 137, 239,\xxb{\bf 307,409,443,613,647,919,953,1021}$
       &40 &4\\
18 & 18 & 95& $37, 73,\xxb{\bf 109,181,397,433,541,577,613,757,829,937  }$   &91 & $[6\cdot10^4]$ \\
   &  0 & 61& $-19,\xxb{\bf -127,-163,-199,-271,-307,-379,-489 }$  &114 & $[10^6]$\\
19 & 19 & 6& $191, 229, 419, 457, 571,\xxb{\bf 647,761,1103,1217,1483,1559}$ &50 &5\\
\midrule
21 & 21 &  & $\xxb{\bf 43,127,211,337,379,421,463,547,631,673,757}$ & &\\
23 & 23 &  & $\xxb{\bf 47,139,277,461,599,691,829,967,1013,1151,1289}$ & &\\
25 & 25 &  & $\xxb{\bf 101,151,251,401,601,701,751,1051,1151,1201}$&&\\
27 & 27 &  & $\xxb{\bf 109,163,271,379,433,487,541,757,811,919,1297}$&&\\
29 & 29 &  & $\xxb{\bf 59,233,349,523,929,1103,1277,1451,1567,1741}$&&\\
33 & 33 &  & $\xxb{\bf 67,199,331,397,463,661,727,859,991,1123,1321}$&&\\
43 & 43 &  & $\xxb{\bf 173,431,947,1033,1291,1549,1721,1979,2237}$&&\\
\midrule
\end{tabular}\label{tab:tab04}}
\end{table}

\begin{table}[!tbh]
  \tbl{The first two period equations for primes $p=ef+1$ with
    $e=21,23,25,27,29,33,39$, and field discriminants $\Delta_e=p^{e-1}$.
    Highlighted are $(\ell_P,\ell_k)$, the number of digits in $D$ and $\Delta_e$.
    The identity $D=\Delta_K$ means $k^2=1$ (See Eq.~(\ref{discri})).
    Large values of $D-\Delta_e$ imply large inessential discriminant divisors.
}
{\begin{tabular}{@{}|c|l|@{}}\toprule
$\Delta_e$ & Coefficients\\
\midrule
$43^{20}$& 1,1,-20,-19,171,153,-816,-680,2380,1820,-4368,-3003,\\
\im(33,33)  &5005,3003,-3432,-1716,1287,495,-220,-55,11,1\\
$127^{20}$& 1,1,-60,-133,1305,4493,-10801,-62425,-2588,380273,489841,-832624,-2307149,\\
\im(104,43) &-540263,3165855,3188668,-157753,-1481414,-380716,205872,50035,-14459\\
\midrule
$47^{22}$& 1,1,-22,-21,210,190,-1140,-969,3876,3060,-8568,-6188,12376,\\
\im(37,37)  &8008,-11440,-6435,6435,3003,-2002,-715,286,66,-12,-1\\
$139^{22}$& 1,1,-66,-147,1630,5648,-16457,-92686,18441,709360,832638,-2239299,-5679764,\\
\im(126,48)&156443,12673530,11318727,-6468097,-14166332,-3186420,5386949,2918745,\\
& -436718,-516219,-63941\\
\midrule
$101^{24}$& 1,1,-48,-43,946,752,-9993,-6962,62052,37341,-234195,-119366,538390,226505,\\
\im(134,49)   &-737819,-249907,571793,151052,-224456,-42136,35494,2561,-1633,-57,19,1\\
$151^{24}$& 1,1,-72,-161,1991,6935,-23789,-131523,59219,1219472,1274267,-5134575,-12138942,\\
\im(150,53) &4263646,39567816,30337248,-42180110,-75903945,-9872226,55689151,\\
&38006340,-5737119,-14814116,-5029783,-190198,111103\\
\midrule
$109^{26}$& 1,1,-52,-47,1128,914,-13369,-9612,95357,60102,-425693,-231576,1201391,553157,-2121177,\\
\im(159,53)   &-810403,2271851,706862,-1399735,-342875,461618,78149,-74294,-4948,4861,-271,-34,1\\
$163^{26}$& 1,1,-78,-175,2388,8354,-33013,-180016,127774,1968453,1782518,-10489594,-23282154,\\
\im(174,58)   &17166283,103371060,63556949,-176991137,-284942103,20494295,355135692,239709074,\\
   &-92044084,-184127684,-70412054,8685910,12922671,3203525,255583\\
\midrule
$59^{28}$& 1,1,-28,-27,351,325,-2600,-2300,12650,10626,-42504,-33649,100947,74613,-170544,-116280,\\
\im(50,50)  &203490,125970,-167960,-92378,92378,43758,-31824,-12376,6188,1820,-560,-105,15,1\\
$233^{28}$& 1,1,-112,-91,5198,3644,-132219,-83238,2053518,1187959,-20553532,-11071128,\\
\im(297,67)   &136460842,69042962,-609473492,-292259011,1836592125,845018358,-3706016039,\\
   &-1661552324,4906886664,2177019390,-4095369839,-1819962089,1998032360,\\
&895362174,-490947342,-221892059,42927079,19524467\\
\midrule
$67^{32}$& 1,1,-32,-31,465,435,-4060,-3654,23751,20475,-98280,-80730,296010,230230,-657800,\\
\im(59,59) &-480700,1081575,735471,-1307504,-817190,1144066,646646,-705432,-352716,293930,\\
  &125970,-77520,-27132,11628,3060,-816,-136,17,1\\
$199^{32}$& 1,1,-96,-217,3795,13403,-74197,-394231,599821,6350170,2832807,-56803105,-104532088,\\
\im(250,74) &244229488,932758015,-5618002,-3890173018,-4529747891,6495127532,18110944809,\\
&4574986912,-26694143816,-29645825157,6037403432,30417132332,15468969217,-6737165737,\\
&-8515927088,-1017730658,1409177433,395068072,-75602260,-21210003,2947097\\
\midrule
$79^{38}$ & 1,1,-38,-37,666,630,-7140,-6545,52360,46376,-278256,-237336,1107568,906192,-3365856,\\
\im(73,73)   &-2629575,7888725,5852925,-14307150,-10015005,20030010,13123110,-21474180,-13037895,\\
&17383860,9657700,-10400600,-5200300,4457400,1961256,-1307504,-490314,\\
&245157,74613,-26334,-5985,1330,190,-20,-1\\
$157^{38}$& 1,1,-76,-71,2556,2222,-50313,-40520,646279,479776,-5720417,-3892342,35931891,22265255,\\
\im(310,84) &-162617513,-91086546,533275855,267613697,-1265136580,-562372122,2154121978,835520674,\\
&-2594978102,-861831376,2164301236,603623323,-1211061590,-280758113,434291871,\\
&84841877,-93357668, -15992102,10935603,1639529,-599706,-67036,11826,272,-49,1\\
\midrule
\end{tabular}\label{tab:tab05}}
\end{table}

Using either polynomial interpolation\cite{bg18} or
systematic computer search\cite{g18},
the number fields underlying these polynomials may be shown to be isomorphic,
with two sets of four transformations interconnecting the polynomials.
The passage from $f(x) \mapsto g(x)$ is accomplished by any of the following
direct transformations
{\small
\begin{eqnarray*}
  D_1 &=& \tfrac{1}{3}( -x^3 - 2x^2 - 18x - 14),\\
  D_2 &=& \tfrac{1}{3}( -x^3 - x^2 - 20x - 6),\\
  D_3 &=& \tfrac{1}{3}(  x^3 + x^2 + 17x + 3),\\
  D_4 &=& \tfrac{1}{3}(  x^3 + 2x^2 + 21x + 14),
\end{eqnarray*}}  
while the inverses, from $g(x) \mapsto f(x)$, are
{\small
\begin{eqnarray*}
  I_1 &=& \tfrac{1}{21}( -2x^3 + 5x^2 - 17x - 7),\\
  I_2 &=& \tfrac{1}{21}( -x^3 - 8x^2 - 19x - 35),\\
  I_3 &=& \tfrac{1}{21}(  x^3 + 8x^2 + 19x + 14),\\
  I_4 &=& \tfrac{1}{21}(  2x^3 - 5x^2 + 17x - 14).
\end{eqnarray*}}

\subsection{How rare are period equations?}

Table \ref{tab:tab04} provides a measure of the relative abundance
and distribution of period equations.
The upper part of the table puts into perspective data from the reference 
database\cite{klu}, while the lower part presents some analogous
results for period equations of larger degrees.
The first column gives polynomial degrees,
the second  records signatures, the third 
gives the number $N$ of polynomials listed in the reference tables,
not necessarily period equations.
In the fourth column, numbers in black inform the basis of field
discriminants of period equations contained in the reference tables.
For instance, among 181 polynomials of degree 10 and signature 10
in the reference table, one finds four period equations whose
discriminants are $41^9, 61^9, 101^9, 181^9$.
Complementing the table, in boldface blue we show bases for the next few
period equations in each sequence.

The fifth column lists the number of digits for the field discriminant
of the last polynomial listed in the reference database.
Thus, for degree 10, the discriminant of the 181th polynomial
with signature $10$ (i.e.~with ten real roots) is a number with 48 digits,
while for signature $0$, the discriminant of  the 79th polynomial contains
22 digits.
As indicated in the rightmost column, such polynomial is the 13th
in the list of signature $0$.
For signature $10$, the corresponding number in the rightmost column is
$[2\cdot10^5]$. Such number is used to indicate that it would take too
much time and resources to establish the sequential order of the 181th
polynomial.
In such cases, numbers in brackets give an estimate of the size of a
prime whose number of digits would be about 48. For instance, the number
of digits of $(2\times 10^5)^9$ is 48.

From the rightmost column of Table  \ref{tab:tab04} one sees clearly
that there is no short supply of period equations.
In particular, it is totally unreasonable to expect any
table to contain them all. From the number of digits listed in the fifth column it
becomes clear that an attempt to include, say, all discriminants for degree 15
would demand a list with no less than 7216 entries.
At the same time, Table  \ref{tab:tab04}
draws attention to how incomplete existing tables  still are, particularly for
polynomials of larger degrees.
For instance, for degree 17, to include all polynomials having discriminants with
up to 40 digits, would require adding just one more polynomial,
the one corresponding to base $307$.
For degree 19, the reference table is already complete for discriminants with
up to 50 digits. The last five lines in Table  \ref{tab:tab04} record some data
for polynomials that we have not found in online tables of number fields.

\subsection{Beyond tabulated polynomials}

Table \ref{tab:tab05}  lists representative pairs of period equations
characterized by
totally real fields for primes $p=ef+1$ where $e=21,23,25,27,29,33$, and $39$.
Coefficients are ordered according to Eq.~(\ref{theta}),
namely for $e=21$ the coefficients are $\alpha_{22}=1$, $\alpha_{21}=11$,
$\alpha_{20}=-55$, etc.

The degrees of the period equations in Table \ref{tab:tab05} go well  beyond what
is presently available in the literature for totally real cyclic fields and
emphasize the easiness  of generating such families systematically.
We computed sequences with varying numbers of period equations, up to $e=100$.
Obviously, such sequences are simply too big to record here, although
they provide significant insight concerning their organization, growth,
as well as minimum discriminants of $\psi_e(x)$.

Note that for the larger degrees, discriminants contain increasingly larger
number of digits, and become harder and harder to factor without
better and dedicated  resources.

\subsection{Orbits and orbital clusters of Pincherle's map $x_{t+1}=2-x_t^2$}
\label{sec:pincherle}

All period equation considered so far had discriminants given by powers of 
single primes.
It is important to mention that $n$-periodic orbits of the quadratic map do not
necessarily have equations of motion defined by $n$ degree polynomials with integer
coefficients as is the case for Eqs.~(\ref{fi}) and (\ref{ficonj}).
Most of the times, orbits appear as {\it orbital clusters\/}
entangling arithmetically together several distinct orbits with the same period.

For instance, for period-$4$,
in the partition generating limit, the limit studied as early as 1920 by
Pincherle\cite{res,pin}, the $3$ individual period-4
orbits  emerge as one single orbit and a cluster formed by two orbits:
\begin{eqnarray*}
  o_{4,1}(x) &=& {x}^{4}+{x}^{3}-4\,{x}^{2}-4\,x+1,\qquad\Delta=3^2\cdot5^3,\\
  c_{4,1}(x) &=& {x}^{8}-{x}^{7}-7\,{x}^{6}+6\,{x}^{5}+15\,{x}^{4}
  -10\,{x}^{3}-10\,{x}^{2}+4\,x+1, \quad\Delta=17^7.
  \end{eqnarray*}
The pair of orbits of the cluster decompose as $c_{4,1}(x)=o_{4,2}(x)\cdot o_{4,3}(x)$, where
{\small
\begin{eqnarray*}
o_{4,2}(x) &=&{x}^{4}-\tfrac{1}{2}(1+\sqrt {17}){x}^{3}-\tfrac{1}{2}(3-\sqrt {17}) {x}^{2}-(2-\sqrt {17}) x-1,
\quad \Delta= 17^2-68\sqrt{17},\\
o_{4,3}(x) &=&{x}^{4}-\tfrac{1}{2}(1-\sqrt {17}){x}^{3}-\tfrac{1}{2}(3+\sqrt {17}) {x}^{2}-(2+\sqrt {17}) x-1,
     \quad\; \Delta= 17^2+68\sqrt{17}.
\end{eqnarray*}}
Remarkably, the single orbit $o_{4,1}(x)$ is not a period equation, while the cluster
$c_{4,1}(x)$ is.
For period-$5$  we obtain
{\small
\begin{eqnarray*}
  o_{5,1}(x) &=& {x}^{5}-{x}^{4}-4\,{x}^{3}+3\,{x}^{2}+3\,x-1,\quad\Delta=11^4,\\
  c_{5,1}(x) &=& {x}^{10}+{x}^{9}-10\,{x}^{8}-10\,{x}^{7}+34\,{x}^{6}
                  +34\,{x}^{5}-43\,{x}^{4}-43\,{x}^{3}\cr
          && +12\,{x}^{2}+12\,x+1, \qquad\quad\Delta=3^5\cdot11^9, \\
  c_{5,2}(x) &=& {x}^{15}-{x}^{14}-14\,{x}^{13}+13\,{x}^{12}+78\,{x}^{11}
               -66\,{x}^{10}-220\,{x}^{9}+165\,{x}^{8}\cr
        && +330\,{x}^{7}-210\,{x}^{6}-252\,{x}^{5}+126\,{x}^{4}+84\,{x}^{3}
-28\,{x}^{2}-8\,x+1. \quad  \Delta=31^4.
\end{eqnarray*}
}%
As before for period-4, $c_{5,1}(x)= o_{5,2}(x)\cdot o_{5,3}(x)$ where
{\small
\begin{eqnarray*}
o_{5,2}(x) &=& {x}^{5}+ \tfrac{1}{2}(1+\sqrt {33}) {x}^{4} -{x}^{3}
  -\tfrac{1}{2}(9+ 3\sqrt {33}) {x}^{2} - (6+ \sqrt {33}) x-1, \quad \Delta=11^2,\\
o_{5,3}(x) &=& {x}^{5}+ \tfrac{1}{2}(1-\sqrt {33}) {x}^{4} -{x}^{3}
  -\tfrac{1}{2}(9- 3\sqrt {33}) {x}^{2} - (6- \sqrt {33}) x-1, \quad \Delta=11^2.
\end{eqnarray*}}
While $o_{5,1}(x)$, the celebrated quintic of Vandermonde\cite{g18}, defines a single
orbit, $c_{5,1}(x)$ and $c_{5,2}(x)$ are entanglements of $2$ and $3$ period-5 orbits,
respectively.
Moreover, $o_{5,1}(x)$, $c_{5,1}(x)$, and $c_{5,2}(x)$ have $k^2=1$, and
$o_{5,1}(x)$ and $c_{5,2}(x)$ are period equations.
As for period-4, individual orbits composing clusters have coefficients given
by complicated algebraic numbers, not integers.
Therefore, orbital clusters may also contain discriminants involving products of
powers of multiple primes.
It is quite challenging to decompose orbital clusters combining more than
two orbits, particularly when they combine an odd number of orbits.
However, the coefficients of such decompositions hide the secretest truth and
most interesting relations among numbers which fix orbital individuality.

Analogously, we find the 18 orbits of period-7  to emerge in three clusters,
with 3, 6, and 9 orbits:
{\small
\begin{eqnarray*}
c_{7,1}(x) &=&   {x}^{21}-{x}^{20}-20\,{x}^{19}+19\,{x}^{18}+171\,{x}^{17}
                -153\,{x}^{16}-816\,{x}^{15}+680\,{x}^{14}\cr
 && \quad\quad +2380\,{x}^{13}-1820\,{x}^{12}-4368\,{x}^{11}+3003\,{x}^{10}
                +5005\,{x}^{9}-3003\,{x}^{8}\cr
 && \quad\quad -3432\,{x}^{7} +1716\,{x}^{6}+1287\,{x}^{5}-495\,{x}^{4}
                -220\,{x}^{3}+55\,{x}^{2}+11\,x-1,\\
c_{7,2}(x) &=& {x}^{42}+{x}^{41}-42\,{x}^{40}-42\,{x}^{39}+  \cdots
           -3267\,{x}^{4}-3267\,{x}^{3}+44\,{x}^{2}+44\,x+1,\\    
c_{7,3}(x) &=&{x}^{63}-{x}^{62}-62\,{x}^{61}+61\,{x}^{60}  +\cdots
             +40920\,{x}^{4}+5456\,{x}^{3}-496\,{x}^{2}-32\,x+1.
\end{eqnarray*}
}%
All three have $k^2=1$ and discriminants $43^{20}$, $3^{21}\cdot43^{41}$ and $127^{62}$,
respectively.
Manifestly, only clusters $c_{7,1}(x)$ and $c_{7,3}(x)$ are period equations.

There is only a quite small number of  non-arithmetically entangled  orbits,
meaning simply that, most of the times, the coefficients of periodic orbits will
be given by more complicated algebraic numbers, not integers.
For periods 9, 10, 11 and 12, the only periodic orbits with integer
coefficients are
{\small
\begin{eqnarray*}
  o_{9,1}(x) &=& {x}^{9}-{x}^{8}-8\,{x}^{7}+7\,{x}^{6}+21\,{x}^{5}-15\,{x}^{4}
                  -20\,{x}^{3}+10\,{x}^{2}+5\,x-1,
                  \quad \Delta=19^8,\\
o_{9,2}(x) &=&{x}^{9}-9\,{x}^{7}+27\,{x}^{5}-30\,{x}^{3}+9\,x-1, \qquad \Delta=3^{22},\\
o_{10,1}(x)&=&{x}^{10}-10\,{x}^{8}+35\,{x}^{6}-{x}^{5}-50\,{x}^{4}+5\,{x}^{3}+25\,{x}^{2}-5\,x-1,
           \qquad \Delta=5^{17},\\
o_{11,1}(x)&=&{x}^{11}-{x}^{10}-10\,{x}^{9}+9\,{x}^{8}+36\,{x}^{7}-28\,{x}^{6}
               -56\,{x}^{5}+35\,{x}^{4}\cr
           && \quad\quad +35\,{x}^{3}-15\,{x}^{2}-6\,x+1, \qquad \Delta=23^{10},\\
o_{12,1}(x)&=&{x}^{12}+{x}^{11}-12\,{x}^{10}-11\,{x}^{9}+54\,{x}^{8}+43\,{x}^{7}-113\,{x}^{6}\cr
           && \quad\quad -71\,{x}^{5}+110\,{x}^{4}+46\,{x}^{3}-40\,{x}^{2}-8\,x+1,
                                      \qquad \Delta= 5^9\cdot7^{10},\\
o_{12,2}(x)&=&{x}^{12}-12\,{x}^{10}+{x}^{9}+54\,{x}^{8}-9\,{x}^{7}-112\,{x}^{6}\cr
           && \quad\quad +27\,{x}^{5}+105\,{x}^{4}-31\,{x}^{3}-36\,{x}^{2}+12\,x+1,
                     \qquad \Delta=  3^{18}\cdot 5^9,\\
o_{12,2}(x)&=&{x}^{12}+{x}^{11}-12\,{x}^{10}-12\,{x}^{9}+53\,{x}^{8}+53\,{x}^{7}\cr
           && \quad\quad -103\,{x}^{6}-103\,{x}^{5}+79\,{x}^{4}+79\,{x}^{3}-12\,{x}^{2}-12\,x+1,
                     \quad \Delta= 3^6\cdot 13^{11}.
\end{eqnarray*}}\noindent
They all have $k^2=1$ and only  $o_{9,1}(x)$ and $o_{11,1}(x)$ are period equations.
The discriminants of $o_{9,1}(x)$ and $o_{9,2}(x)$ are the first and second smallest for
cyclic equations of degree nine, while $o_{10,1}(x)$ has the third smallest
and $o_{11,1}(x)$  the smallest possible discriminant for cyclic polynomials
of degrees 10 and 11, respectively\cite{klu}.
Of the 335 period-12 orbits, only the three above have integer coefficients.
They are not period equations, but are
the triplet of cyclic polynomials with minimum discriminants.
The 630 period-13 orbits emerge as three rather big polynomials, of degrees
1365, 2730, and 4095, conglomerating  105, 210,  and 315 orbits, respectively.
The coefficients of the individual orbits must involve algebraic numbers
with exquisite symmetry properties that would be interesting to study,
despite the challenge of the task.
For polynomial maps, an exact equation giving the total number of periodic
orbits as a function of the period is given in Ref.~\cite{count}.

Among orbits and orbital clusters of the quadratic map one finds the
startling phenomenon of {\it period inheritance}\cite{epl}.
A detailed discussion of orbits and clusters for the quadratic map
will be presented elsewhere.

\section{Conclusions and outlook}

Motivated by the remarkable fact that several periodic orbits and orbital clusters
of the quadratic map coincide with period equations, 
this paper reported a number of properties of period equations uncovered by
computing large sets of them, and consolidating trends observed.

It was found that period equations may be systematically generated and enumerated,
with no omissions,  for primes of the form $p=ef+1$.
This fact allows one to recognize and to fix some gaps in tables of number
fields currently available in the literature.
It also makes clear that, due to the abundance of period equations, there is 
no hope of ever producing ``complete'' tables. Fortunately, however,
period equations are not difficult to generate when needed, using
currently available computer algebra systems.
Maybe future version of such type of software will incorporate intrinsic
functions for this purpose.

The design of an efficient routine for the systematic determination of classes
of solutions ended up disclosing exact theoretical expressions, conjectures,
which seem hard to come by theoretically and which are now ready to be challenged
by traditional demonstrations.
For instance, we found a simple and  general closed-form expression,
Eq.~(\ref{conjecture}), for the field discriminant of cyclotomic period equations.
As shown, such expression grants direct access to the so-called inessential
discriminant divisors\cite{ha64} buried in  conventional polynomial discriminants
and normally quite difficult to determine.
Equation (\ref{conjecture}) provides an easy criterion to sort out equations with
either $k^2=1$ or $k^2\neq1$, sets that we find to contain an unbounded quantity
of equations and emerging intertwined with a quite irregular distribution of
magnitudes and signs.
Additional analytical results are reported in Section \ref{sec:linear},
in particular by Eq.~({\ref{geral}), 
and in Sections \ref{sec:tab} and \ref{sec:ide}.

As is the case for $\varphi_1(x)$ and $\varphi_2(x)$,
note that the {\it branch ambiguity} of the square root signs in $o_{4,2}(x)$
and $o_{4,3}(x)$, as well as in $o_{5,2}(x)$
and $o_{5,3}(x)$, make such orbits to be only {\sl formally} well defined.
In fact, to represent unambiguous orbits, such forms still depend on fixing
the branch for the square root that they contain.
By suitable branch choice, the formally ambiguous expression $o_{4,2}(x)$ may be
``projected'' into anyone of the two branch-fixed orbits.
The same is valid for the ambiguous $o_{4,3}(x)$ that
may be also selected to represent anyone of the two branch-fixed orbits.
The existence of root-ambiguity before fixing branches is a simple arithmetical
consequence of the multivaluedness of numbers in the roots.

The systematic generation of period equations allows one to enumerate unambiguously
classes of number fields.
In some sense, such enumeration resembles somewhat the arithmetic order
discovered to exist among unordered binary labels associated with the symbolic
dynamics of the quadratic map\cite{eg06}.
The identification of period equations as periodic orbits and clusters
of the quadratic map lends hope that, eventually,  it may be possible to
disclose analytically the regular processes underlying the organization of
bifurcation cascades observed so frequently in physical models.
A promising application is to detect and classify orbital interdependencies
in classical dynamics\cite{epl}.
An open question is to understand why some orbital equations are {\it not}
period equations, while some clusters are.
A further enticing open question is to determine if, as for period equations,
other discriminant regularities abundantly present in number field databases are
associated with additional families of polynomials yet to be discovered.

\medskip\par\noindent
{\bf Note added (October 31, 2019):}
While searching for references to Eqs.~(\ref{conjecture}) and (\ref{geral}),
with the help of the internet and kind leads and feedback provided by Profs.
G.E.~Andrews, B.C.~Berndt,  R.J.~Evans, K.~Gy\H{o}ry, F.~Lemmermeyer,
W.~Narkiewicz, A.~Schinzel, A.~Ware, H.C.~Williams, and K.S.~Williams,
it was possible to uncover the following facts.

Prof.~Evans pointed out that, up to sign, Eq.~(\ref{conjecture}) was given
by Neto et al.~\cite{nil}.  
By Galois theory, there is only one possible subfield $K$ and, accordingly,
we identify $[K:\mathbb Q]=e$.
The field $K$ is generated over $\mathbb Q$ by the $e$ roots of the period
equation $\psi_e(x)$  in Eq.~(\ref{theta}).
Our Eqs.~(\ref{conjecture}) and (\ref{conj2}) agree with the magnitude of
$\Delta_e$ reported by Neto et al., and, in addition, they provide the
proper signs for all cases.

Prof.~Narkiewicz pointed out that our Eq.~(\ref{conjecture}) is correct and
follows from the conductor-discriminant formula, see, e.g., Theorem 3.11 in
the book {\sl Introduction to Cyclotomic Fields}\cite{was}.
He also mentions that Gurak \cite{gugu} presented a procedure to
compute the beginning coefficients of the minimal polynomials of the
period equations. It was not verified if Gurak's results lead or not to our
Eq.~(\ref{geral}).

It was not yet possible to locate exact reference to Eq.~(\ref{geral})
in the literature.
However, using cyclotomic numbers and other results from the book
{\sl Gauss and Jacobi Sums}\cite{bew},
Prof.~K.S.~Williams 
kindly sent us a general proof that Eq.~(\ref{geral}) is indeed correct,
as well as an expression for the  missing case of totally complex fields
with signature $(0,e/2)$, namely
{\small
\begin{equation}
      \alpha_2 = \frac{e+p-1}{2e} = \tfrac{1}{2}(f+1).   \label{ken}
\end{equation}}
This coefficient matches exactly all our computational  data.
The author expresses his gratitude to all persons involved for their
generous  contributions.

\section*{Acknowledgments}
This work was supported by the Max-Planck Institute for the Physics of
Complex Systems, Dresden, in the framework of the Advanced Study Group on
{\sl Forecasting with Lyapunov vectors}.
The author was supported by CNPq, Brazil.





\end{document}